\documentclass[aps,eqsecnum,preprint,floats,epsf,epsfig,nofootinbib]{revtex4}
\usepackage{epsfig}
\usepackage{graphicx}
\usepackage{extarrows}
\usepackage{subfigure}

\begin{document}
\def\be{\begin{eqnarray}}
\def\en{\end{eqnarray}}
\def\non{\nonumber}
\def\la{\langle}
\def\ra{\rangle}
\def\A{{\cal A}}
\def\B{{\cal B}}
\def\c{{\cal C}}
\def\d{{\cal D}}
\def\e{{\cal E}}
\def\p{{\cal P}}
\def\t{{\cal T}}
\def\nc{N_c^{\rm eff}}
\def\CP{{\it CP}~}
\def\CPP{{\it CP}}
\def\acp{{\cal A}_{C\!P}}
\def\vp{\varepsilon}
\def\drho{\bar\rho}
\def\deta{\bar\eta}
\def\vma{{_{V-A}}}
\def\vpa{{_{V+A}}}
\def\J{{J/\psi}}
\def\ov{\overline}
\def\Lqcd{{\Lambda_{\rm QCD}}}
\def\pr{{ Phys. Rev.}~}
\def\prl{{ Phys. Rev. Lett.}~}
\def\pl{{ Phys. Lett.}~}
\def\np{{ Nucl. Phys.}~}
\def\zp{{ Z. Phys.}~}
\def\lsim{ {\ \lower-1.2pt\vbox{\hbox{\rlap{$<$}\lower5pt\vbox{\hbox{$\sim$}
}}}\ } }
\def\gsim{ {\ \lower-1.2pt\vbox{\hbox{\rlap{$>$}\lower5pt\vbox{\hbox{$\sim$}
}}}\ } }


\vskip 1.0 cm

\centerline{\large\bf {\it CP} violation
 in the interference between $\rho(770)^0$ and $S$-wave in }
\centerline{\large \bf $B^\pm\to \pi^+ \pi^-\pi^\pm$ decays and its implication for {\it CP} asymmetry
in $B^\pm\to \rho^0\pi^\pm$}
\bigskip  \medskip
\centerline{\bf Hai-Yang Cheng}
\medskip
\centerline{Institute of Physics, Academia Sinica}
\centerline{Taipei, Taiwan 115, Republic of China}
\medskip
\bigskip

\bigskip
\bigskip
\centerline{\bf Abstract}
\bigskip

\small
The decay amplitude analyses of $B^\pm\to \pi^+ \pi^-\pi^\pm$ decays in the Dalitz plot performed by LHCb indicate that \CP asymmetry for the dominant quasi-two-body decay $B^\pm\to\rho(770)^0\pi^\pm$ was found to be consistent with zero in all three approaches for the $S$-wave component and that  \CPP-violation effects related to the interference between the $\rho(770)^0$ resonance and the $S$-wave were clearly observed. We show that the nearly vanishing \CP violation in $B^\pm\to\rho^0\pi^\pm$ is understandable within the framework of QCD factorization. There are two $1/m_b$ power corrections, one from penguin annihilation and the other from hard spectator interactions. They contribute destructively to $\acp(B^\pm\to\rho^0\pi^\pm)$ to render it compatible with zero, in contrast to the sizable negative \CP asymmetry predicted in most of the existing models.
We next show that the measured interference pattern between the $\rho$ and $S$-wave contributions in the low $m_{\pi^+\pi^-}$ region separated by the sign of value of $\cos\theta$ with $\theta$ being the angle between the two same charged pions measured in the rest frame of the $\rho$ resonance can be explained in terms of the smallness of $\acp(B^+\to\rho^0\pi^+)$ and the interference between $\rho(770)$ and $\sigma/f_0(500)$. If \CP asymmetry in $B^\pm\to\rho^0\pi^\pm$ is not negligible as predicted in many existing models, the observed interference pattern will be destroyed. We conclude that the experimental observation of the interference pattern between $P$- and $S$-waves in the low-$m_{\pi^+\pi^-}$ region between 0.5 and 1.0 GeV is consistent with a nearly vanishing \CP violation in $B^\pm\to\rho^0\pi^\pm$.

\pagebreak

\section{Introduction}

Recently, the full amplitude analysis of $B^\pm\to \pi^+\pi^-\pi^\pm$ in the Dalitz plot has been performed by LHCb \cite{Aaij:3pi_1,Aaij:3pi_2}.
In this analysis, the $S$-wave component of $B^\pm\to \pi^+\pi^-\pi^\pm$ was studied using three different approaches: the isobar model, the $K$-matrix model and a quasi-model-independent binned approach. In the isobar model, the $S$-wave amplitude was presented by LHCb as a coherent sum of the $\sigma$ (or $f_0(500)$) meson contribution and a $\pi\pi\leftrightarrow K\ov K$ rescattering amplitude in the mass range $1.0<m_{\pi^+\pi^-}<1.5$ GeV. The fit fraction of the $S$-wave is about 25\% and predominated by the $\sigma$ resonance.

\begin{figure}[t]
\begin{center}
\includegraphics[width=0.70\textwidth]{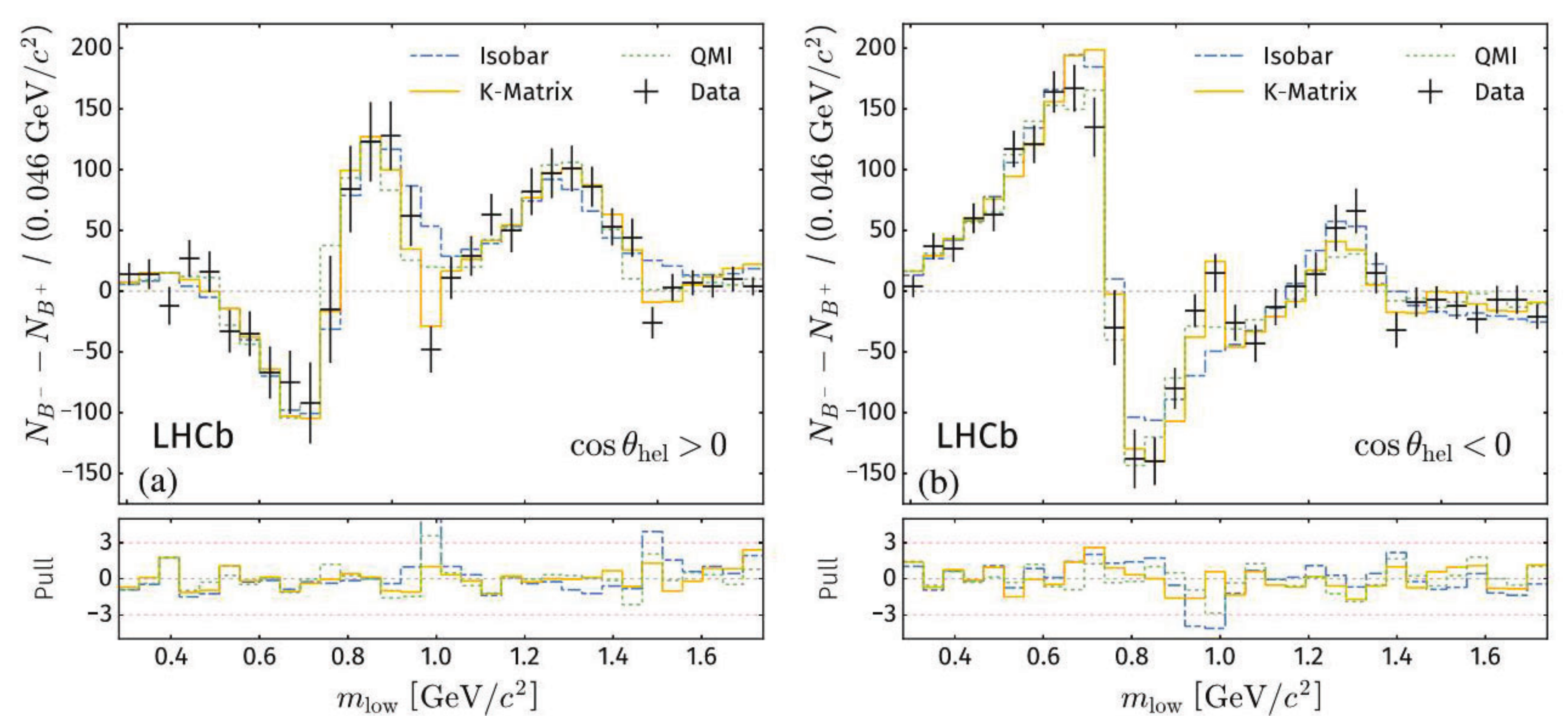}
\vspace{0.1cm}
\caption{The difference of $N_{B^-}$ and $N_{B^+}$,  the number of $B^-$ and $B^+$ events respectively, for $B^\pm\to \pi^+\pi^-\pi^\pm$ measured in the low-$m_{\pi^+\pi^-}$ region for (a) $\cos\theta_{\rm hel}>0$ and (b) $\cos\theta_{\rm hel}<0$ with the helicity angle $\theta_{\rm hel}$ being defined in Fig. \ref{fig:theta}. This plot is taken from Ref. \cite{Aaij:3pi_2}.}
\label{fig:rho-sigma}
\end{center}
\end{figure}

\begin{figure}[t]
\begin{center}
\includegraphics[width=0.35\textwidth]{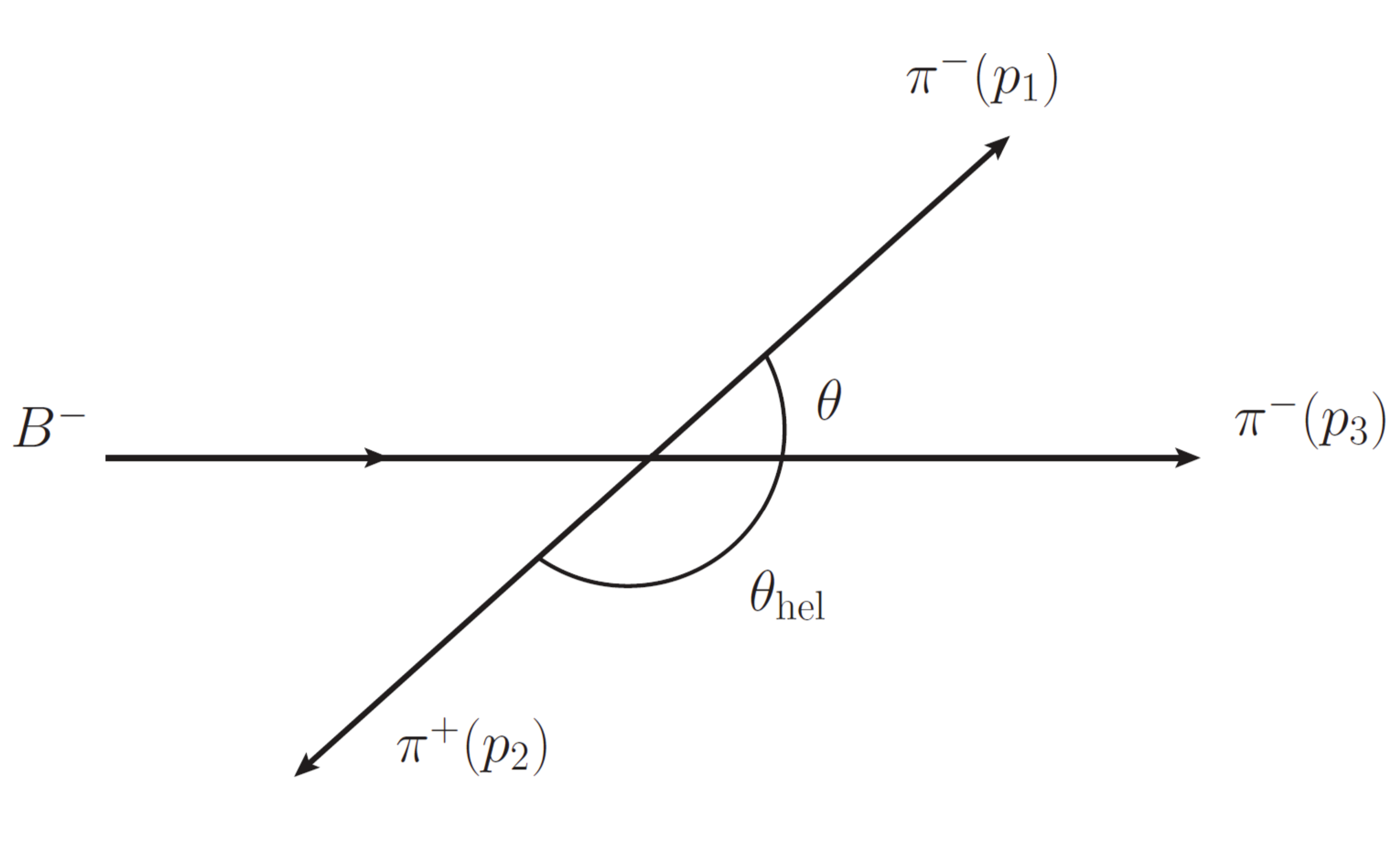}
\vspace{0.0cm}
\caption{The angle $\theta$ between the momenta of the two $\pi^-$ pions measured in the rest frame of the dipion system in the decay $B^-\to \pi^-(p_1)\pi^+(p_2)\pi^-(p_3)$. It is related to the helicity angle $\theta_{\rm hel}$ defined by the LHCb through the relation $\theta+\theta_{\rm hel}=\pi$.} \label{fig:theta}
\end{center}
\end{figure}

In the low invariant-masss $m(\pi^+\pi^-)_{\rm low}$ region, \CP asymmetries were clearly seen in $B^\pm\to \pi^+\pi^-\pi^\pm$ decays in the following places: (i)
a significant \CP violation of 15\%  in $B^\pm\to\sigma\pi^\pm$ implied in the isobar model,
(ii) a  large \CP asymmetry of order $45\%$ in the rescattering amplitude,
(iii) a \CP violation of 40\% in the mode with the tensor resonance $f_2(1270)$, and
(iv)  large \CPP-violating effects related to the interference between the $P$- and $S$-wave contributions with the significance exceeding $25\sigma$ in all the $S$-wave models.   More precisely, the interference between $S$- and $P$-waves is clearly visible in Fig. \ref{fig:rho-sigma}  where the data are separated by the sign of the value of $\cos\theta_{\rm hel}$ with $\theta_{\rm hel}$ being the helicity angle, evaluated in the $\pi^+\pi^-$ rest frame, between the pion with opposite charge to the $B$ and the third pion from the $B$ decay (see Fig. \ref{fig:theta}). \footnote{The same effect also can be seen in Fig. 3 of Ref. \cite{Aaij:3pi_1} where \CP asymmetries are plotted as a function of $\cos\theta_{\rm hel}$ in regions below and above the $\rho(770)$ resonance pole.}
In contrast, \CP asymmetry for the dominant quasi-two-body decay mode $B^\pm\to\rho(770)^0\pi^\pm$ was found by LHCb to  be compatible with zero in all three $S$-wave approaches.

We notice two salient features in Fig. \ref{fig:rho-sigma}: (i)
\CP violation in the $\rho(770)$ region is proportional to $(m_{\rm low}^2-m_\rho^2)\cos\theta_{\rm hel}$ as the sign of \CP asymmetry is flipped below and above the $\rho(770)$ peak and from negative $\cos\theta_{\rm hel}$ to positive $\cos\theta_{\rm hel}$.  Hence, \CP asymmetry in this region arises from the interference between the $\rho(770)$ and $S$-wave contributions because the angular distribution of $\rho\to\pi^+\pi^-$ is proportional to $\cos\theta_{\rm hel}$, while it is a constant for $S\to \pi^+\pi^-$.
The change of asymmetry below and above the $\rho(770)$ peak implies a strong phase arising from the $\rho$ line shape.
(ii) The height of the peak and the depth of the valley near the $\rho$ mass region are similar though the former is slightly larger than the latter.

To describe the observed interference pattern, we consider the interference between $\rho(770)$ and $\sigma(500)$ as the $S$-wave is predominated by the $\sigma$ resonance in the isobar model. Writing
\be \label{eq:Apm}
A_\pm\equiv A^\rho_\pm+A^\sigma_\pm=c_\pm^\rho F_\rho^{\rm BW}\cos\theta+ c_\pm^\sigma F_\sigma^{\rm BW}
\en
for the $B^+$ and $B^-$ decays, where $A^{\rho(\sigma)}_\pm\equiv A(\B^\pm\to\rho^0(\sigma) \pi^\pm\to \pi^-(p_1)\pi^+(p_2)\pi^\pm(p_3))$,
$F_\rho^{\rm BW}$ and $F_\sigma^{\rm BW}$ are the Breit-Wigner propagators of the $\rho(770)$ and $\sigma$, respectively, for example,
\be
  F_\rho^{\rm BW}(s_{12})={1\over s_{12}-m_\rho^2+ im_\rho\Gamma_\rho}.
\en
We use the Breit-Wigner line shape for the purpose of illustration. In realistic calculations given in Sec. IV below, we shall utilize the Gounaris-Sakurai model to describe the line shape of the $\rho(770)$ meson and the pole model for the $\sigma$ resonance. In Eq. (\ref{eq:Apm}), we have followed Ref. \cite{Bediaga:2015} to define the quantity $\theta$ being the angle between the two same charged pions measured in the rest frame of the dipion system (i.e. the resonance). This angle is related to the helicity angle $\theta_{\rm hel}$ defined by LHCb \cite{Aaij:3pi_2} through the relation $\theta_{\rm hel}+\theta=\pi$ (see Fig. \ref{fig:theta}). Hence, $\cos\theta_{\rm hel}=-\cos\theta$.

It follows that \CP asymmetry has the expression
\be \label{eq:modelCP}
\A_{C\!P} &\propto& |A_-|^2-|A_+|^2 \non \\
&=&
(|c_-^\rho|^2-|c_+^\rho|^2)|F_\rho^{\rm BW}(s_{12})|^2 \cos\theta^2+  (|c_-^\sigma|^2-|c_+^\sigma|^2)|F_\sigma^{\rm BW}(s_{12})|^2   \non \\
&+& 2\,|F_\rho^{\rm BW}(s_{12})|^2 |F_\sigma^{\rm
BW}(s_{12})|^2 \cos\theta \Bigg\{ {\rm Re}(c_-^{*\rho}c_-^\sigma-c_+^{\rho}c_+^{*\sigma})\Big[(s_{12}-m_\rho^2)(s_{12}-m_\sigma^2) \non \\
&+& (m_\rho\Gamma_\rho)(m_\sigma\Gamma_\sigma)\Big]
+{\rm Im}(c_-^{*\rho}c_-^\sigma+c_+^{\rho}c_+^{*\sigma}) \Big[m_\rho\Gamma_\rho(s_{12}-m_\sigma^2)- m_\sigma\Gamma_\sigma(s_{12}-m_\rho^2)\Big]\Bigg\}.
\en
The terms $(s-m_{\rho(\sigma)}^2)$ and $m_{\rho(\sigma)}\Gamma_{\rho(\sigma)}$ arise from the real and imaginary parts, respectively, of the Breit-Wigner propagator $F_{\rho(\sigma)}^{\rm BW}(s)$.
The observed interference pattern shown in Fig. \ref{fig:rho-sigma} in the $\rho$ mass region indicates that it is mainly governed by the $(s_{12}-m_\rho^2)\cos\theta$ term. This requires that \CP violation induced by the $\rho$ meson is negligible; that is $|c_-^\rho|^2\approx |c_+^\rho|^2$, and that the pattern is not significantly affected by the \CP asymmetry induced by $\sigma$ in the region of interest. A nearly vanishing \CP violation originating from $\rho$ is consistent with the fact that the \CPP-violating effect cancels when integrating over the angle. In other words, the observed interference pattern between $\rho$ and $S$-wave is consistent with the smallness of \CP violation in $B^\pm\to\rho^0\pi^\pm$.

It is a  long-standing puzzle in regard to the \CP asymmetry in  $B^\pm\to\rho^0\pi^\pm$.  The existing theoretical
predictions based on QCD factorization (QCDF) \cite{CC:Bud,Sun:2014tfa}, perturbative QCD (pQCD) \cite{LiYa:2016,Chai:2022kmq}, soft-collinear effective theory (SCET) \cite{Wang:2008rk}, topological diagram approach (TDA) \cite{Cheng:TDA,Zhou:2016jkv}  all lead to a negative \CP asymmetry for $B^\pm\to\rho^0\pi^\pm$ except the pQCD calculation in \cite{Chai:2022kmq}, ranging from $-7\%$ to $-45\%$ (see Eq. (\ref{eq:Theory_rhoCP}) below).

The purpose of this work is twofold. First, we would like to point out that the nearly vanishing \CP violation in $B^\pm\to\rho^0\pi^\pm$ is understandable in the framework of QCDF.
Second, we will present a study of \CP asymmetries in the low-$m_{\pi\pi}$ region between 0.5 and 1.0 GeV and show that the interference pattern depicted in Fig. 1 can be explained provided that  \CP violation in $B^\pm\to\rho^0\pi^\pm$ is negligible.

This paper is organized as follows. In Sec. II we study \CP violation in $B^\pm\to\rho^0\pi^\pm$ decays in the framework of QCDF. We then proceed to the $B^\pm\to \sigma\pi^\pm$ decays in Sec. III. In Sec. IV we consider the three-body decays $B^\pm\to\pi^-\pi^+\pi^\pm$  mediated by the $\rho(770)$ and $\sigma(500)$ resonances and study their interference. We summarize our results in Section V.

\section{$B^\pm\to \rho^0\pi^\pm$ decays}
\CP asymmetry in the quasi-two-body decay  $B^\pm\to \rho^0\pi^\pm$  was found by LHCb to  be consistent with zero in all three $S$-wave approaches \cite{Aaij:3pi_1,Aaij:3pi_2}.
In the isobar model,
\be
\A_{CP}(B^+\to\rho^0\pi^+)=(0.7\pm1.1\pm0.6\pm1.5)\%, \qquad {\rm LHCb~2020.}
\en
Recently, this asymmetry was measured by LHCb based on a method that does not require full amplitude analysis \cite{LHCb:BtoPV}. The result is
\be
\A_{CP}(B^+\to\rho^0\pi^+)=(-0.4\pm1.7)\%, \qquad {\rm LHCb~2022.}
\en
Predictions of \CP violation (in \%) in various approaches are summarized as follows:
\be \label{eq:Theory_rhoCP}
\A_{CP}(B^+\to\rho^0\pi^+)= \left\{ \begin{array}
{r l l }
{\rm QCDF}: & -9.8^{+3.4+11.4}_{-2.6-10.2}~[4], & -6.7^{+0.2+3.2}_{-0.2-3.7}~[5] \\
{\rm pQCD}: & -27.5^{+2.5}_{-3.3}\pm1.7~[6], &  +14.9^{+0.4+0.5}_{-0.4-0.6}~[7] \\
{\rm SCET}: &  -19.2^{+15.5+1.7}_{-13.4-1.9}~[8], & -10.8^{+13.1+0.9}_{-12.7-0.7}~[8] \\
{\rm TDA}:  & -23.9\pm 8.4~[9],  & -45\pm4~[10]
\end{array} \right.
\en
Hence, except the pQCD calculation in \cite{Chai:2022kmq} all the exisiting models predict
a negative \CP asymmetry ranging from $-7\%$ to $-45\%$.

It has been claimed in \cite{Bediaga:2016} that in $B\to PV$ decays with $m_V<1$ GeV, \CP asymmetry induced from a short-distance mechanism is suppressed by the $C\!PT$ constraint.
$C\!PT$ theorem generally implies the same lifetimes for both particle and antiparticle.
When partial widths are summed over, the total width of the particle and its antiparticle should be the same. Final-state interactions are responsible for  distributing the \CP asymmetry among the different conjugate decay channels. In the three-body $B$ decays, the ``2+1" approximation is usually assumed so that the resonances produced in heavy meson decays do not interact with the third particle. Within the ``2+1" approximation, the observation of a large negative \CP asymmetry of order $-27\%$ for the $B^0\to K^*(892)^+\pi^-$ decay with $K^*(892)^+\to K^0\pi^+$ \cite{LHCb:Kstpi} is compensated by a similarly large positive \CP asymmetry in the channel  $B^0\to K^*(892)^+\pi^-$ followed by $K^*(892)^+\to K^+\pi^0$; that is \cite{Bediaga:2020qxg},
\be
\A_{CP}(B^0\to K^*(892)^+\pi^-\to K^+\pi^0\pi^-)\approx -\A_{CP}(B^0\to K^*(892)^+\pi^-\to K^0\pi^+\pi^-).
\en
This can be tested once \CP asymmetry in $B^0\to K^0\pi^+\pi^-$ is measured in the future.
Likewise, $\A_{CP}(B^+\to \rho^0\pi^+\to \pi^+\pi^-\pi^+)$ is expected to vanish as the $\rho^0$ cannot decay into two neutral pions. By the same token, one will also expect a vanishing
$\A_{CP}(B^+\to \rho^0 K^+\to \pi^+\pi^- K^+)$. However, this is in contradiction to the experimental observation of $\A_{C\!P}(B^+\to\rho^0 K^+)=0.150\pm0.019$ by LHCb recently \cite{LHCb:BtoPV}. In this case, one may argue that there are other possibilities that can produce \CP violation, for example, a three-body rescattering including the third particle \cite{Bediaga:2016}.


In QCDF, the decay amplitude of $B^-\to\rho^0\pi^-$ is given by \cite{BN}
\be \label{eq:Amprhopi}
A(B^-\to\rho^0\pi^-) &=&  \frac{G_F}{2}\sum_{p=u,c}\lambda_p^{(d)}\Bigg\{ \left[\delta_{pu}(a_2-\beta_2)-a_4^p-r_\chi^\rho a_6^p+{3\over 2}(a_7^p+a_9^p)+{1\over 2}(a_{10}^p+r_\chi^\rho a_8^p) \right. \non \\
&&  -\beta_3^p-\beta^p_{\rm 3,EW}\Big]_{\pi\rho} X^{(B^-\pi,\rho)}
+ \Big[\delta_{pu}(a_1+\beta_2)+a_4^p-r_\chi^\pi a_6^p+a_{10}^p-r_\chi^\pi a_8^p
  \non \\
&& \left.+\beta_3^p+\beta^p_{\rm 3,EW}\right]_{\rho\pi}X^{(B^-\rho,\pi)},
\en
where $\lambda_p^{(d)}=V_{pb}V_{pd}^*$, the chiral factors $r_\chi^{\pi,\rho}$ are given by
\be \label{eq:rchi}
 r_\chi^\pi(\mu)={2m_\pi^2\over m_b(\mu)(m_u+m_d)(\mu)},  \qquad r_\chi^\rho(\mu) = \frac{2m_\rho}{m_b(\mu)}\,\frac{f_\rho^\perp(\mu)}{f_\rho} \,,
\en
and the factorizable matrix elements read
\be
X^{(B^-\pi,\rho)} =2f_\rho m_B p_c F_1^{B\pi}(m_\rho^2), \qquad
X^{(B^-\rho,\pi)} =2f_\pi m_B p_c A_0^{B\rho}(m_\pi^2),
\en
with $p_c$ being the c.m. momentum. Here we have followed \cite{BSW} for the definition of form factors.
In Eq. (\ref{eq:Amprhopi}), the flavor operators $a_i^{p}$
have the expressions \cite{BBNS,BN}
\be \label{eq:ai}
  a_i^{p}(M_1M_2) =
 c_i+{c_{i\pm1}\over N_c}
  + {c_{i\pm1}\over N_c}\,{C_F\alpha_s\over
 4\pi}\Big[V_i(M_2)+{4\pi^2\over N_c}H_i(M_1M_2)\Big]+P_i^{p}(M_2),
\en
where $i=1,\cdots,10$,  the upper (lower) signs apply when $i$ is
odd (even), $c_i$ are the Wilson coefficients,
$C_F=(N_c^2-1)/(2N_c)$ with $N_c=3$. The
quantities $V_i(M_2)$ account for vertex corrections,
$H_i(M_1M_2)$ for hard spectator interactions with a hard gluon
exchange between the emitted meson and the spectator quark of the
$B$ meson and $P_i(M_2)$ for penguin contractions.
The order of the arguments of the $a_i^p(M_1M_2)$ and
$\beta_i(M_1M_2)$ coefficients to be mentioned below is dictated by the subscript $M_1M_2$.

It is known that the leading-order $1/m_b$ predictions of QCDF encounter several major difficulties: (i) the predicted branching fractions for penguin-dominated $B\to PP,VP,VV$ decays are systematically below the measurements, (ii) the predicted rates for color-suppressed tree-dominated decays $B^0\to\pi^0\pi^0,\rho^0\pi^0$ are too small,  and  (iii) direct \CP asymmetries for $\bar B^0\to K^-\pi^+, K^{*-}\pi^+, \pi^+\pi^-$, $B^-\to K^-\rho^0$ and $\bar B_s\to K^+\pi^-$,$\cdots$ etc., disagree with experiment in signs \cite{Cheng:2009,CC:Bud}. Hence, it is necessary to introduce $1/m_b$ power corrections to resolve above-mentioned difficulties. To solve the problems with (i) and (iii) in QCDF, power corrections to the penguin amplitudes are described by the penguin annihilation characterized by the parameters $\beta_i^p$ (see Eq. (19) of Ref. \cite{BN} for the definition), for example, $\beta_{2,3}^p$ and $\beta^p_{\rm 3,EW}$ given in Eq. (\ref{eq:Amprhopi}). Take the decay $\bar B^0\to K^-\pi^+$ as an example. In the heavy quark limit, the calculated branching fraction is too small by around 35\% and the predicted \CP violation is positive in sign (see e.g. Table IV of Ref. \cite{CC:Bud}). However, the first evidence of direct \CP asymmetry in $B$ decays, namely, $\A_{CP}(\bar B^0\to K^-\pi^+)=-0.133\pm0.031$ observed in 2004 \cite{BaBar:2004gyj} is negative. Power corrections from penguin annihilation solve the rate deficit and the sign problems at one stroke, see again Table IV of Ref. \cite{CC:Bud}.

As pointed out in Refs. \cite{Cheng:2009,CC:Bud}, sign flip of \CP asymmetries in penguin-dominated due to the presence of power corrections from penguin annihilation will cause another \CP puzzle: (iv) signs of $\acp$ in $B^-\to K^-\pi^0,~K^-\eta,~\pi^-\eta,\cdots$ will also get reversed in such a way that they disagree with experiment. It turns out that the difficulties with (ii) and (iv) can be resolved by invoking power corrections to hard spectator interactions. To see this, hard spectator interactions for $i=1-4,9,10$ have the expression \cite{BN}
\begin{eqnarray}\label{eq:hardspec}
  H_i(M_1 M_2)= {if_B f_{M_1} f_{M_2} \over X^{(\overline{B} M_1,
  M_2)}}\,{m_B\over\lambda_B} \int^1_0 d x \int_0^1 d y \,
 \Bigg( \frac{\Phi_{M_2}(x) \Phi_{M_1}(y)}{\bar x\bar y} + r_\chi^{M_1}
  \frac{\Phi_{M_2}(x) \Phi_{m_1} (y) }{\bar x y}\Bigg),
 \hspace{0.5cm}
 \end{eqnarray}
with $\bar x=1-x$ and $\bar y=1-y$. Subleading $1/m_b$ power corrections stem from the twist-3 amplitude $\Phi_m$.
Hard spectator contributions to $a_i$ are most prominent for $a_2$ as the Wilson coefficient $c_1$ is of order unity
\be \label{eq:a2}
a_2(M_1M_2) = c_2+{c_1\over N_c} + {c_1\over N_c}\,{C_F\alpha_s\over
 4\pi}\Big[V_2(M_2)+{4\pi^2\over N_c}H_2(M_1M_2)\Big].
\en
This explains why hard spectator interactions are helpful to account for the  observed rates of $B^0\to \pi^0\pi^0$ and $\rho^0\pi^0$.

In the QCD factorization approach, power corrections often involve endpoint divergences, for example, the second term in Eq. (\ref{eq:hardspec}) is proportional to the endpoint divergence $X\equiv\int^1_0 dy/y$.
We shall follow \cite{BBNS} to model the endpoint divergence in the penguin annihilation and hard spectator scattering diagrams as
\be \label{eq:XA}
 X_A = \ln\left({m_B\over \Lambda_h}\right)(1+\rho_A e^{i\phi_A}), \qquad
 X_H = \ln\left({m_B\over \Lambda_h}\right)(1+\rho_H e^{i\phi_H}),
\en
with $\Lambda_h$ being a typical hadronic scale of 0.5 GeV. We add the superscripts `$V\!P$' and `$PV$' to distinguish penguin annihilation effects in $B\to V\!P$ and $B\to PV$ decays \cite{BN}:
\be
&& A_1^i\approx -A_2^i\approx 6\pi\alpha_s\left[3\left(X_A^{VP}-4+{\pi^2\over 3}\right)+r_\chi^V r_\chi^P\Big((X_A^{VP})^2-2X_A^{VP}\Big)\right], \non \\
&& A_3^i\approx  6\pi\alpha_s\left[-3r_\chi^V\left((X_A^{VP})^2-2X_A^{VP}+4-{\pi^2\over 3}\right)+r_\chi^P \left((X_A^{VP})^2-2X_A^{VP}+{\pi^2\over 3}\right)\right], \non \\
&& A_3^f\approx  6\pi\alpha_s\left[3r_\chi^V(2X_A^{VP}-1)(2-X_A^{VP})-r_\chi^P \Big(2(X_A^{VP})^2-X_A^{VP}\Big)\right],
\en
for $M_1M_2=V\!P$  and
\be
&& A_1^i\approx -A_2^i\approx 6\pi\alpha_s\left[3\left(X_A^{PV}-4+{\pi^2\over 3}\right)+r_\chi^V r_\chi^P\Big((X_A^{PV})^2-2X_A^{PV}\Big)\right], \non \\
&& A_3^i\approx  6\pi\alpha_s\left[-3r_\chi^P\left((X_A^{PV})^2-2X_A^{PV}+4-{\pi^2\over 3}\right)+r_\chi^V \left((X_A^{PV})^2-2X_A^{PV}+{\pi^2\over 3}\right)\right], \non \\
&& A_3^f\approx  6\pi\alpha_s\left[-3r_\chi^P(2X_A^{PV}-1)(2-X_A^{PV})+r_\chi^V \Big(2(X_A^{PV})^2-X_A^{PV}\Big)\right],
\en
for $M_1M_2=PV$, where the superscripts `$i$' and `$f$' refer to gluon emission from the initial and final-state quarks, respectively, the subscripts 1 for the Dirac structure $(V-A)\otimes (V-A)$, 2 for $(V-A)\otimes(V+A)$ and 3 for $-2(S-P)\otimes(S+P)$.

The two unknown parameters $\rho_A$ and $\phi_A$ were fitted to the data of $B\to PP, V\!P$ and $PV$ decays. The  values of $\rho_A$ and $\phi_A$ are given, for example, in Table III of Ref. \cite{CC:Bud}, where the results are very similar to the so-called ``S4 scenario" presented in \cite{BN}. However, the measurement of the pure annihilation process $B_s^0\to\pi^+\pi^-$ by  CDF \cite{CDF:Bspipi} and LHCb \cite{LHCb:Bspipi} with the world average $\B(B_s\to\pi^+\pi^-)=(0.70\pm0.10)\times 10^{-6}$ \cite{PDG} is much higher than the QCDF prediction of $(0.26^{+0.00+0.10}_{-0.00-0.09})\times 10^{-6}$ \cite{Cheng:Bs}.
Since this mode proceeds through the penguin-annihilation amplitudes $A_1^i$ and $A_2^i$, it
has been advocated that $\rho_A$'s associated with the gluon emission from the initial and final-state quarks are different; that is, $\rho_A^i\neq \rho_A^f$ and that a much large $\rho_A^i$ of order 3 can accommodate the data of $B_s^0\to\pi^+\pi^-$ \cite{Zhu:2011mm,Wang:2013fya}.

Nevertheless, a recent study of $B_s^0\to \pi^+\pi^-$ in QCDF up to the NLL order given by \cite{Lu:2022fgz}
\be
A(B_s^0\to \pi^+\pi^-)={G_F\over \sqrt{2}}\sum_{p=u,c}\lambda_p^{(s)}(if_B f_\pi^2) 4\pi\alpha_s{C_F\over N_c^2}\left(T^{p,(0)}+{\alpha_s\over 4\pi}T^{p,(1)}+\cdots \right)
\en
shows that the real part of ~$T^{c,(0)}$ is enhanced significantly, while the imaginary part gets considerable cancellation. More precisely, $T^{c,(0)}=-0.25+2.16i$ while
$\alpha_s/(4\pi)T^{c,(1)}=-1.82-2.40i$\, at $\mu=m_b$. Indeed, the weak annihilation diagram with the gluon emission from the initial-state quark is calculable to the leading power in $1/m_b$ expansion. It yields $(\rho_A^i,\phi_A^i)_{P\!P}=\{(0.97, -97^\circ),~(1.17, -97^\circ),~(1.34,-97^\circ)\}$ for $\lambda_B=\{200, 350, 500\}$ MeV, respectively \cite{Lu:2022fgz}.  This implies that $(\rho^i_A)_{P\!P}$ is of order unity rather than 3. Moreover, the new measurement of another pure annihilation process $\B(B^0\to K^+K^-)=(7.80\pm1.27\pm0.84)\times 10^{-8}$ by LHCb \cite{LHCb:K+K-} is strongly disfavored with a large $(\rho^i_A)_{P\!P}$ as the predicted $\B(B^0\to K^+K^-)$ will be too large by a factor of 3 $\sim 4$. Therefore, we will not distinguish between $\rho_A^i$ and $\rho_A^f$.

For $B\to PV$ decays, $(\rho_H,\phi_H)$, $(\rho_A, \phi_A)_{_{V\!P}}$ and $(\rho_A, \phi_A)_{_{PV}}$ are treated as free parameters.
In this work, we shall take
\be \label{eq:rhoA}
(\rho_A, \phi_A)_{_{PV}}=(0.82, -37^\circ), \qquad (\rho_A, \phi_A)_{_{V\!P}}=(0.68, -52^\circ),
\en
for calculations, where only the central values are listed here.   The parameters $(\rho_A, \phi_A)_{_{PV}}$ are inferred from $B\to \pi K^*$ decays, while $(\rho_A, \phi_A)_{_{V\!P}}$ from $B\to \rho K$ ones. In the $P\!P$ sector, we have $(\rho_A, \phi_A)_{_{P\!P}}=(1.05, -40^\circ)$.
For \CP violation in the $V\!P$ sector, we find
\be
\A_{CP}(\bar B^0\to K^{*-}\pi^+) = -(17.3^{+0.9+3.7}_{-1.3+3.2})\%, \qquad
\A_{CP}(B^-\to \rho^0 K^-) = (28.7^{+21.5+2.5}_{-20.5-3.3})\%,
\en
while the experimental measurements are $-0.271\pm0.044$~\cite{LHCb:Kstpi} and
$0.150\pm0.019$~\cite{LHCb:BtoPV}, respectively.

\begin{table}[t]
\caption{The branching fraction and \CP asymmetry of $B^-\to \rho^0\pi^-$ within the QCDF approach. Experimental data are taken from \cite{PDG}. The theoretical errors correspond to the uncertainties due to the variation of (i) Gegenbauer moments, decay constants,  form factors, the strange quark mass, and (ii) $\rho_{A,H}$, $\phi_{A,H}$, respectively. In (ii) we assign an error of $\pm0.4$ to $\rho$ and $\pm 4^\circ$ to $\phi$.}
\label{tab:rhopi_theory}
\begin{center}
\begin{tabular}{ l r l} \hline \hline
 $\B(10^{-6})$ & ~~$\A_{C\!P}(\%)$ &  ~~~Comments~~ \\
\hline
 $8.3\pm1.2$ & $0.7\pm1.9$ & ~~~Experiment \\
 \hline
 $8.3^{+1.9+0.0}_{-0.9-0.0}$ & ~~$6.6^{+0.5+0.0}_{-0.8-0.0}$ & ~~~(1) Heavy~quark~limit \\
 $8.5^{+1.9+0.2}_{-0.9-0.2}$ & $-11.8^{+0.9+3.6}_{-0.7-3.9}$ & ~~~(2) $\rho_{H}=0$ and $\phi_{H}=0$~ with~$\rho_A$ and $\phi_A$ given by Eq. (\ref{eq:rhoA})  \\
 $7.5^{+1.5+0.1}_{-0.7-0.2}$ & $15.6^{+1.3+1.2}_{-1.1-1.5}$ &  ~~~(3) $\rho_H=3.15,~\phi_H=-113^\circ$, $\rho_A=0$, $\phi_A=0$ \\
 $7.7^{+1.5+0.2}_{-0.7-0.2}$ & $-0.5^{+3.9+3.5}_{-2.6-3.8}$ & ~~~(4) $\rho_H=3.15$, ~$\phi_H=-113^\circ$, $\rho_A$ and $\phi_A$ given by Eq. (\ref{eq:rhoA}) \\
\hline \hline
\end{tabular}
\end{center}
\end{table}

As noticed in passing, it is necessary to introduce the power corrections from hard spectator interactions characterized by $\rho_H$ and $\phi_H$ to not only account for the rate deficit of the color-suppressed tree-dominated modes $B^0\to \pi^0\pi^0$ and $\rho^0\pi^0$ but also resolve the
\CP puzzles with some of the penguin-dominated modes such as  $B^-\to K^-\eta$,  $B^0\to K^{*0}\eta$ and $B^-\to K^-\pi^0$, where the last one is related to the so-called $\Delta \A_{K\pi}$ puzzle with $\Delta \A_{K\pi}\equiv \A_{CP}(B^-\to K^-\pi^0)-\A_{CP}(\bar B^0\to K^-\pi^+)$.
In the absence of hard spectator interactions, \CP asymmetries of $B^-\to K^-\eta$,  $B^0\to K^{*0}\eta$, $B^-\to K^-\pi^0$ and $\Delta \A_{K\pi}$ are calculated to be 0.045, 0.010, $-0.035$ and 0.022, respectively, while the corresponding measured values are $-0.37\pm0.08$, $0.19\pm0.05$, $0.030\pm0.013$ and $0.115\pm0.014$ \cite{PDG}. One needs hard spectator interactions to flip the signs of $\A_{CP}(B^-\to K^-\eta)$, $\A_{CP}(B^-\to K^-\pi^0)$ and enhance $\A_{CP}(B^0\to K^{*0}\eta)$.
Specifically,  we shall use
\be
(\rho_H, \phi_H)_{_{PV,VP}}=(3.15\,, -113^\circ), \qquad (\rho_H, \phi_H)_{_{P\!P}}=(4.0\,, -70^\circ).
\en
to accommodate the above-mentioned data. Notice that $\rho_H$ is much larger than $\rho_A$.

We are now ready to compute the branching fraction and \CP asymmetry for $B^-\to \rho^0\pi^-$, see Table \ref{tab:rhopi_theory}. In the heavy quark limit, its \CP asymmetry is positive with a magnitude of order 0.07\,. We then turn on power corrections induced from penguin annihilation. It is clear that the sign of $\acp(\rho^0\pi^-)$ is flipped and in the meantime its magnitude is enhanced. We next switch on $1/m_b$ corrections from hard spectator interactions and turn off $\rho_A$ and $\phi_A$. It is evident  that $\acp(\rho^0\pi^-)$ will be enhanced from ${\cal O}(0.07)$ to ${\cal O}(0.16)$ in the presence of hard spectator effects alone.
If the heavy quark limit of $\acp(\rho^0\pi^-)$ is considered as a benchmark, hard spectator interactions will push it up  further, whereas penguin annihilation will pull it to the opposite direction. Therefore, the nearly vanishing  $\acp(\rho^0\pi^-)$ arises from two $1/m_b$ power corrections which contribute destructively.

At the first sight, it appears from Table \ref{tab:rhopi_theory} that the calculated branching fraction of order $7.7\times 10^{-6}$ is slightly smaller than the experimental value of $(8.3\pm1.2)\times 10^{-6}$ \cite{PDG}. In the PDG \cite{PDG}, the branching fraction of $B^-\to \rho^0\pi^-$ is extracted from the measurement of $B^-\to \rho^0\pi^-\to \pi^+\pi^-\pi^-$ through the factorization relation
\be
\B(B^-\to \rho^0\pi^-\to \pi^+\pi^-\pi^-)=\B(B^-\to \rho^0\pi^-)\B(\rho^0\to \pi^+\pi^-)=\B(B^-\to \rho^0\pi^-),
\en
which is valid only in the narrow width approximation (NWA). Since $\rho(770)$ is broad with a width of 149 MeV, it is necessary to consider the finite-width effect. The $\eta_R$ parameters for various resonances produced in three-body $B$ decays have been evaluated in
\cite{Cheng:2020mna,Cheng:2020iwk}
\be
\B(B\to RP)=\eta_R\B(B\to RP)_{\rm NWA}=\eta_R{\B(B\to RP_3\to P_1P_2P_3)_{\rm expt}\over
\B(R\to P_1P_2)_{\rm expt}}
~.
\en
The parameter $\eta_\rho$ depends on the line shape of the $\rho(770)$ meson and it has been calculated in QCDF \cite{Cheng:2020mna,Cheng:2020iwk}. It was found $\eta_\rho^{GS}=0.93$ in the Gounaris-Sakurai (GS) model~\cite{Gounaris:1968mw} to be discussed below for the $\rho$ line shape adapted by both LHCb \cite{Aaij:3pi_1,Aaij:3pi_2} and BaBar \cite{BaBarpipipi} in the analysis of $B^\pm\to\pi^+\pi^-\pi^\pm$ decays. It follows that the PDG value of $\B(B^-\to\rho\pi^-)=(8.3\pm1.2)\times 10^{-6}$  should be corrected to $(7.7\pm1.1)\times 10^{-6}$ after including finite-width effects.


\section{$B^\pm\to \sigma\pi^\pm$ decays}

The decay $B^-\to \sigma/f_0(500)\pi^-$ has been studied in Ref. \cite{Cheng:2020hyj}.
Its decay amplitude  has a similar expression as $B^-\to f_0(980)\pi^-$ and reads
\be \label{eq:Ampsigmapi}
A(B^- \to \sigma \pi^- ) &=&
 \frac{G_F}{\sqrt{2}}\sum_{p=u,c}\lambda_p^{(d)}
 \Bigg\{ \left[a_1 \delta_{pu}+a^p_4+a_{10}^p-(a^p_6+a^p_8) r_\chi^\pi \right]_{\sigma\pi} X^{(B\sigma,\pi)} \non \\
 &+&
 \left[a_2\delta_{pu} +2(a_3^p+a_5^p)+{1\over 2}(a_7^p+a_9^p)+a_4^p-{1\over 2}a_{10}^p-(a_6^p-{1\over 2}a_8^p)\bar r^\sigma_\chi\right]_{\pi\sigma}\ov X^{(B\pi,\sigma)}\non \\
 &-& f_Bf_\pi\bar f_{\sigma}^u\bigg[\delta_{pu}b_2(\pi\sigma)+ b_3(\pi\sigma)
 + b_{\rm 3,EW}(\pi\sigma) +(\pi\sigma\to \sigma\pi) \bigg] \Bigg\},
\en
where the factorizable matrix elements are given by
\be
X^{(B\sigma,\pi)}=-f_\pi F_0^{B\sigma^u}(m_\pi^2)(m_B^2-m_\sigma^2), \qquad
\ov X^{(B\pi,\sigma)}=\bar f_\sigma^u F_0^{B\pi}(m_\sigma^2)(m_B^2-m_\pi^2),
\en
with $\bar r_\chi^{\sigma}(\mu)=2m_{\sigma}/m_b(\mu)$.
The superscript $u$ in the scalar decay constant $\bar f_\sigma^u$ and the form factor $F^{B\sigma^u}$ refers to the $u$ quark component of the $\sigma$.
The scale-dependent scalar decay constant is defined by
$\la\sigma |\bar uu|0\ra=m_\sigma \bar f_\sigma^u$ \cite{CCY:SP}.  We follow~\cite{Cheng:2020hyj} to take $\bar f_\sigma^u=350$ MeV at $\mu=1$ GeV and $F_0^{B\sigma^u}(0)=0.25$\,.

Numerical values of the flavor operators $a_i^p(\sigma\pi)$ and $a_i^p(\pi\sigma)$ at the scale $\mu=m_b(m_b)$ and penguin annihilation characterized by the parameter $\beta^p$ are shown
in Table IV of Ref. \cite{Cheng:2020hyj}. It follows that the flavor operators $a_i^p(\sigma\pi)$ and $a_i^p(\pi\sigma)$ are very different except for $a_{6,8}^p$ as the former do not receive factorizable contributions. As a consequence, $a_1(\sigma\pi)\approx 1\gg a_1(\pi\sigma)$.

Numerically, we obtain
\be \label{eq:CPsigma}
\B(B^-\to\sigma\pi^-)=(5.15^{+1.31+0.86}_{-1.16-1.29})\times 10^{-6}, \qquad
\A_{C\!P}(B^-\to\sigma\pi^-)=(15.10^{+0.31+~8.36}_{-0.30-11.38})\%,
\en
where use of the Wolfenstein parameters updated with $A=0.8132$, $\lambda=0.22500$, $\bar \rho=0.1566$ and $\bar \eta=0.3475$ \cite{CKMfitter} has been made.
Theoretical uncertainties come from (i) the Gegenbauer moments $B_{1,3}$, the scalar meson decay constants, the heavy-to-light form factors and
the strange quark mass, and (ii) the power corrections due to weak annihilation and hard spectator interactions, respectively.
The calculated \CP asymmetry agrees well with the LHCb measurement analyzed in the isobar model \cite{Aaij:3pi_1,Aaij:3pi_2}
\be
\A_{C\!P}(B^-\to\sigma\pi^-)=(16.0\pm1.7\pm2.2)\%.
\en

\section{Interference between $\rho(770)$ and $\sigma(500)$ resonances}

In this section we consider the three-body $B^-\to\pi^-\pi^+\pi^-$ decays mediated by the $\rho(770)$ and $\sigma(500)$ resonances and study their interference.
For the three-body decay amplitude
$A^{\rho}_-\equiv A(\B^-\to\rho^0(770) \pi^-\to \pi^-(p_1)\pi^+(p_2)\pi^-(p_3))$, factorization leads to the expression~\cite{Cheng:2020ipp}
\be \label{eq:Arho}
A^{\rho}_- = -g^{\rho\to \pi^+\pi^-}\,F(s_{12},m_{\rho})T_{\rho}^{\rm GS}(s_{12})2q\cos\theta\, \tilde A(B^-\to \rho\pi^-) +(s_{12}\leftrightarrow s_{23}),
\en
where $q$ given by
\be \label{eq:3momentum}
q=|\vec{p}_1|=|\vec{p}_2|={1\over 2}\sqrt{s_{12}-4m_\pi^2},
\en
is the momenta of $\pi^-(p_1)$ and $\pi^+(p_2)$ in the rest frame of the $\rho$ with the invariant mass $m_{12}=\sqrt{s_{12}}$,
$\tilde A(B^-\to \rho\pi^-)$ has the same expression as $A(B^-\to \rho\pi^-)$ given in Eq. (\ref{eq:Amprhopi}) except for the replacement of $X^{(B^-\pi,\rho)}$ and $X^{(B^-\rho,\pi)}$ by
\be
\tilde X^{(B^-\pi,\rho)} &=& 2f_\rho m_B \tilde p_c F_1^{B\pi}(s_{12}), \non \\
\tilde X^{(B^-\rho,\pi)} &=& 2 f_\pi m_B\tilde p_c\Big[
A_0^{B\rho}(m_\pi^2)
+  {1\over 2m_\rho}\left(m_B-m_{\rho}-{m_B^2-s_{12}\over m_B+m_{\rho}}\right)A_2^{B\rho}(m_\pi^2)\Big],
\en
respectively, with \footnote{The momentum $|\vec{p}_3|$ is related to $\tilde p_c$ by the relation $|\vec{p}_3|=(m_B/m_{12})\tilde p_c$.}
\be
\tilde p_c=\left({(m_B^2-m_\pi^2-s_{12})^2-4s_{12}m_\pi^2 \over 4m_B^2}\right)^{1/2}.
\en
being the c.m. momentum of $\pi^-(p_3)$ and the $\rho(m_{12})$  in the $B$ rest frame.
It is easily seen that when $s_{12}\to m_\rho^2$, $\tilde A(B^-\to \rho\pi^-)$ is reduced to the QCDF amplitude $A(B^-\to \rho\pi^-)$  given by Eq. (\ref{eq:Amprhopi}).

When $\rho$ is off the mass shell, especially when $s_{12}$ is approaching the upper bound of $(m_B-m_\pi)^2$, it is necessary to account for the off-shell effect. For this purpose, we shall follow~\cite{Cheng:FSI} to introduce a form factor $F(s,m_R)$ parameterized as
\be \label{eq:FF for coupling}
F(s,m_R)=\left( {\Lambda^2+m_R^2 \over \Lambda^2+s}\right)^n,
\en
with the cutoff $\Lambda$ not far from the resonance,
\be
\Lambda=m_R+\beta\Lambda_{\rm QCD},
\en
where the parameter $\beta$ is expected to be of order unity. We shall use $n=1$, $\Lambda_{\rm QCD}=250$ MeV and $\beta=1.0\pm0.2$ in subsequent calculations.

In Eq. (\ref{eq:Arho}), $T_\rho^{\rm GS}$ is a description of the broad $\rho(770)$ resonance in the Gounaris-Sakurai (GS) model~\cite{Gounaris:1968mw}.  It was employed by both BaBar~\cite{BaBarpipipi} and LHCb~\cite{Aaij:3pi_1,Aaij:3pi_2} Collaborations in their analyses of the $\rho(770)$ resonance in the $B^-\to \pi^-\pi^+\pi^-$ decay.
The GS line shape for $\rho(770)$ is given by
\be
T_\rho^{\rm GS}(s)={ 1+D \, \Gamma_\rho^0/m_\rho \over s-m^2_{\rho}-f(s)+im_{\rho}\Gamma_{\rho}(s)},
\label{eq: T GS}
\en
where
\be
\Gamma_{\rho}(s)=\Gamma_{\rho}^0\left( {q\over q_0}\right)^3
{m_{\rho}\over \sqrt{s}} {X^2_1(q)\over X^2_1(q_0)},
\en
with $\Gamma_\rho^0$ being the nominal total $\rho$ width with $\Gamma_\rho^0=\Gamma_\rho(m_\rho^2)$ and
the Blatt-Weisskopf barrier factor given by $X_1^2(z)=1/[(zr_{\rm BW})^2+1]$ and $r_{\rm BW}\approx 4.0~{\rm GeV}^{-1}$.
The quantity $q_0$ is the value of $q$ when $m_{12}$ is equal to the pole mass $m_\rho$.
In this model, the real part of the pion-pion scattering amplitude with an intermediate $\rho$ exchange calculated from the dispersion relation is taken into account by the $f(s)$ term in the propagator of $T_\rho^{\rm GS}(s)$.
Explicitly,
\be \label{eq:f(s)}
f(s)=\Gamma_\rho^0{m_\rho^2\over q_0^3}\left[ q^2[h(\sqrt{s})-h(m_\rho)]+(m_\rho^2-s)q_0^2\left.{dh\over ds}\right\vert_{m_\rho}\right],
\en
and
\be
h(s)={2\over \pi}{q\over \sqrt{s}}\ln\left( {\sqrt{s}+2q\over 2m_\pi}\right), \qquad
\left.{dh\over ds}\right\vert_{m_\rho}=h(m_\rho)\left[ {1\over 8q_0^2}-{1\over 2m_\rho^2}\right]+{1\over 2\pi m_\rho^2}.
\en
The constant parameter $D$ is given by
\be
D={3\over \pi}\,{m_\pi^2\over q_0^2}\ln \left( {m_\rho+2q_0\over 2m_\pi} \right)+{m_\rho\over 2\pi q_0}-{m_\pi^2 m_\rho\over \pi q^3_0}.
\en

Likewise, for the decay amplitude $A^\sigma_-\equiv A(B^-\to\sigma\pi^-\to \pi^+\pi^-\pi^-)$, factorization leads to~\cite{Cheng:2020ipp}
\be \label{eq:Asigma}
A^{\sigma}_-=
 g^{\sigma\to \pi^+\pi^-} F(s_{12},m_\sigma)\,T_\sigma(s_{12})\tilde A(B^-\to \sigma \pi^-)+ (s_{12}\leftrightarrow s_{23}),
\en
where $\tilde A(B^-\to \sigma\pi^-)$ has the same expression as $A(B^-\to \sigma\pi^-)$ given in Eq. (\ref{eq:Ampsigmapi}) with the replacement
\be
X^{(B\sigma,\pi)} &\to& \tilde X^{(B\sigma,\pi)}=-f_\pi (m_B^2-s_{12})F_0^{B\sigma^u}(m_\pi^2), \non \\
X^{(B\pi,\sigma)} &\to& \tilde X^{(B\pi,\sigma)}=\bar f_\sigma^u (m_B^2-m_\pi^2)F_0^{B\pi}(s_{12}).
\en
For the $\sigma$ line shape we follow the LHCb Collaboration~\cite{Aaij:3pi_2} to use the simple pole description
\be
T_\sigma(s)={1\over s-s_\sigma}={1\over s-m_\sigma^2+\Gamma_\sigma^2(s)/4+im_\sigma\Gamma_\sigma(s)},
\en
with
\be
\Gamma_{\sigma}(s)=\Gamma_{\sigma}^0\left( {q\over q_0}\right)
{m_{\sigma}\over \sqrt{s}},
\en
and
\be
\sqrt{s_\sigma}=m_\sigma-i\Gamma_\sigma/2=(563\pm 10)-i(350\pm13)\,{\rm MeV},
\en
obtained from the isobar description of the $\pi^+\pi^-$ $S$-wave to fit the $B^-\to\pi^-\pi^+\pi^-$ decay data~\cite{Aaij:3pi_2}.

Strong coupling constants $g^{\rho\to \pi^+\pi^+}$ in Eq. (\ref{eq:Arho}) and $g^{\sigma\to\pi^+\pi^-}$ in Eq. (\ref{eq:Asigma}) are determined from the measured widths through the relations
\be \label{eq:partialwidth_1}
 \Gamma_{S\to P_1P_2}={p_c\over 8\pi m_S^2}g_{S\to P_1P_2}^2,\qquad
 \Gamma_{V\to P_1P_2}={p_c^3\over 6\pi m_V^2}g_{V\to P_1P_2}^2.
\en
Numerically,
\be
|g^{\rho\to \pi^+\pi^+}|=6.00\,, \qquad |g^{\sigma\to\pi^+\pi^-}|=3.90\,{\rm GeV}.
\en

In the LHCb experiment, \CP violation induced by the interference between $P$- and $S$-waves is measured in the low-$m_{\pi^+\pi^-}$ region where the data are separated by the sign of $\cos\theta$ or $\cos\theta_{\rm hel}$, see Fig. \ref{fig:rho-sigma}.
The $\cos\theta$ term can be expressed  as a function of $s_{12}$ and $s_{23}$
\be
\cos\theta=a(s_{12})s_{23}+b(s_{12}),
\en
with \cite{Bediaga:2015}
\be
a(s) &=& {1\over (s-4m_\pi^2)^{1/2}\left({(m_B^2-m_\pi^2-s)^2\over 4s} - m_\pi^2\right)^{1/2}}, \non \\
b(s) &=& -{m_B^2+3m_\pi^2-s\over 2(s-4m_\pi^2)^{1/2}\left({(m_B^2-m_\pi^2-s)^2\over 4s} - m_\pi^2\right)^{1/2}}.
\en
Note that the cosine of the angle $\theta$ with the values $-1$, 0 and 1  corresponds to $(s_{23})_{\rm min}=-(1+b)/a$, $s_{23}=-b/a$ and
$(s_{23})_{\rm max}=(1-b)/a$, respectively.

\begin{figure}[tbp]
\centering
\subfigure[]
{
  \includegraphics[width=6.5cm]{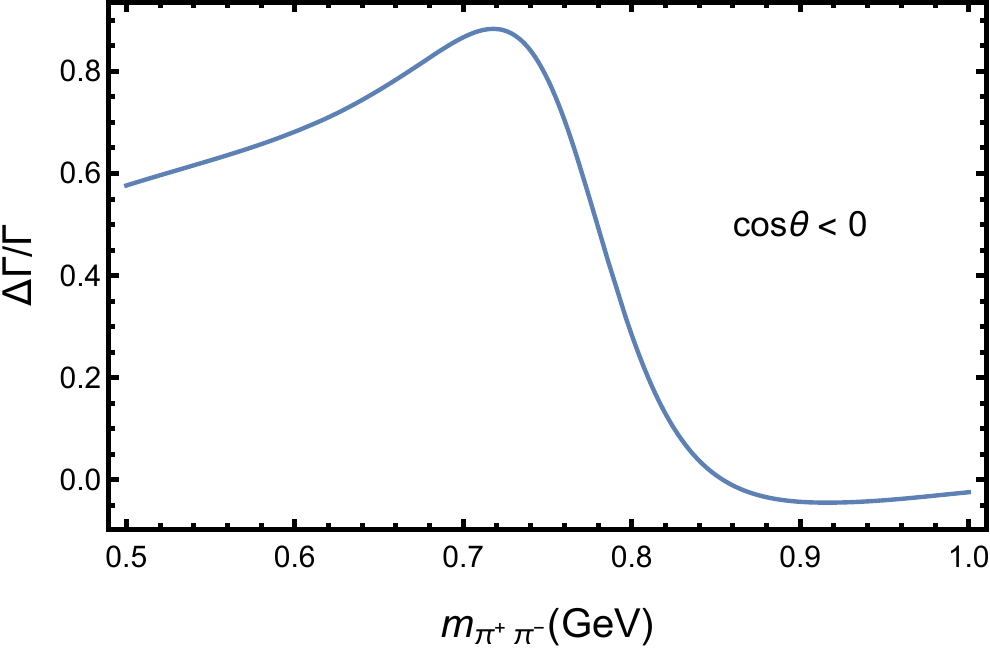}
}
\hspace{0.2cm}
\subfigure[]
{
  \includegraphics[width=6.5cm]{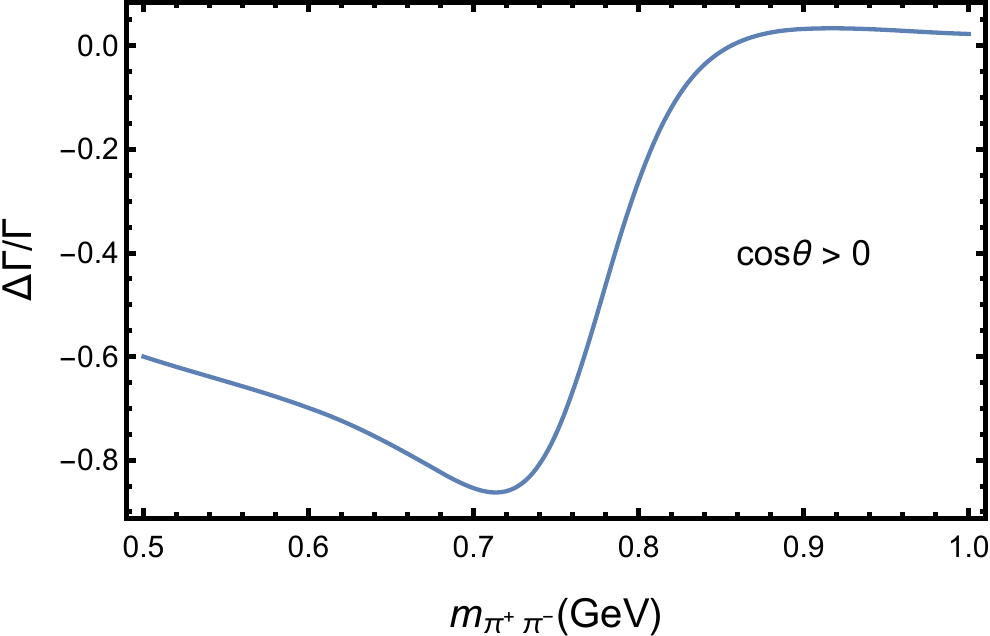}
}
\caption{ Rate asymmetries $\Delta\Gamma$ in units of $\Gamma=1/\tau(B^\pm)$ for $B^\pm\to \pi^\pm\pi^+\pi^-$ in the low-$m_{\pi^+\pi^-}$ region induced by the $\rho(770)$ and $\sigma$ interference for (a) $\cos\theta<0$ ($\cos\theta_{\rm hel}>0$)  and (b)  $\cos\theta>0$ ($\cos\theta_{\rm hel}<0$). The relative phases between $A^\sigma_\pm$ and $A^\rho_\pm$ are set to zero.
}
\label{fig: CPrho-sigma}
\end{figure}

The 3-body decay rates involving $\rho$, $\sigma$ and their interference in the negative and positive $\cos\theta$ regions are given by
\be \label{eq:Rate}
\Gamma^{\rho-\sigma}_\pm(s_{12})^{(\cos\theta<0)} &=& {1\over (2\pi)^3 32m_B^3}\,{G_F^2\over 2}\,{1\over 2}\int_{(s_{23})_{\rm min}}^{-b/a}\Big( |A^\rho_\pm|^2+|A^\sigma_\pm|^2+2\big[{\rm Re}(A^\rho_\pm){\rm Re}(A^\sigma_\pm)  \non \\
&& \hspace{4.2cm} +{\rm Im}(A^\rho_\pm){\rm Im}(A^\sigma_\pm)\big]\Big)ds_{23}
  ~~~{\rm for}~\cos\theta<0,   \non \\
\Gamma^{\rho-\sigma}_\pm(s_{12})^{(\cos\theta>0)} &=& {1\over (2\pi)^3 32m_B^3}\,{G_F^2\over 2}\,{1\over 2}\int^{(s_{23})_{\rm max}}_{-b/a}\Big( |A^\rho_\pm|^2+|A^\sigma_\pm|^2+2\big[{\rm Re}(A^\rho_\pm){\rm Re}(A^\sigma_\pm)  \non \\
&& \hspace{4.2cm} +{\rm Im}(A^\rho_\pm){\rm Im}(A^\sigma_\pm)\big]\Big)ds_{23}
  ~~~{\rm for}~\cos\theta>0,
\en
where the identical particle effect has been taken care of by the factor of 1/2, and the amplitudes
$A^\rho_-$ and $A^\sigma_-$ are given by Eqs. (\ref{eq:Arho}) and (\ref{eq:Asigma}), respectively. In the above equation, the $\rho(770)$ and $\sigma$ interference is depicted by the terms $[{\rm Re}(A^\rho_\pm){\rm Re}(A^\sigma_\pm)+{\rm Im}(A^\rho_\pm){\rm Im}(A^\sigma_\pm)]$.
The rate difference $\Delta \Gamma^{\rho-\sigma}\equiv \Gamma^{\rho-\sigma}_--\Gamma^{\rho-\sigma}_+$ due to the $\rho(770)$ and $\sigma$ interference is shown in Figs. \ref{fig: CPrho-sigma}(a) and \ref{fig: CPrho-sigma}(b)
for $\cos\theta<0$ (or $\cos\theta_{\rm hel}>0$) and $\cos\theta>0$ (or $\cos\theta_{\rm hel}<0$), respectively. It is obvious that they do not resemble the data displayed in Fig. \ref{fig:rho-sigma} at all. For example, the shapes are quite different from that of the data and the characteristic feature of $(s_{12}-m_\rho^2)\cos\theta$ is absent. \footnote{In our previous work \cite{Cheng:2020ipp}, Figs. 5(a) and 5(b) were obtained by setting $\beta$'s, the parameters describing weak annihilation, to be zero. The relative phases between $A^\sigma_\pm$ and $A^\rho_\pm$ were also not taken into account.}

\begin{figure}[tbp]
\centering
\subfigure[]{
  \includegraphics[width=6.5cm]{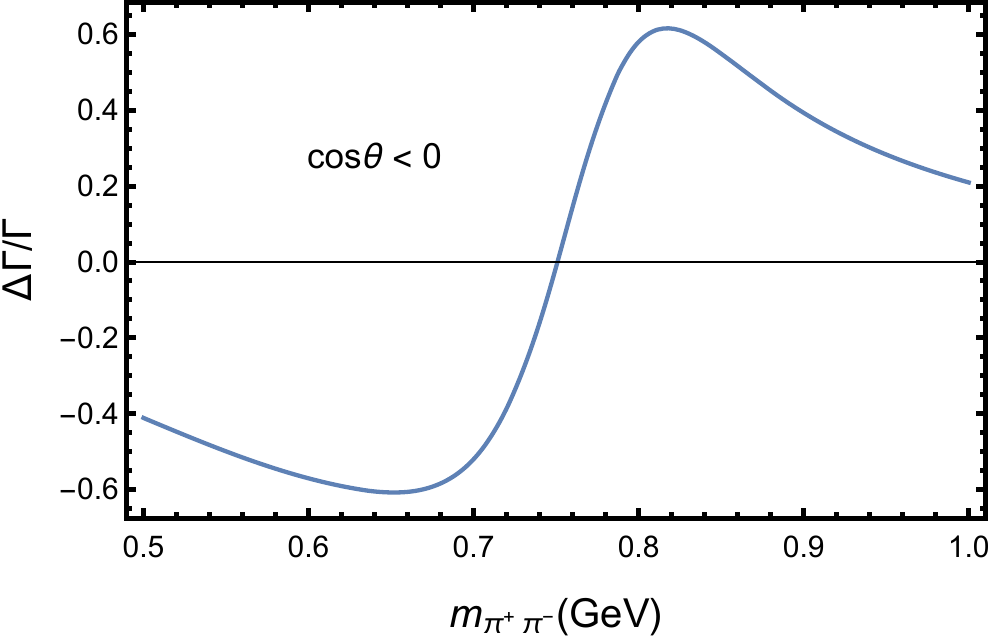}
}
\hspace{0.5cm}
\subfigure[]{
  \includegraphics[width=6.5cm]{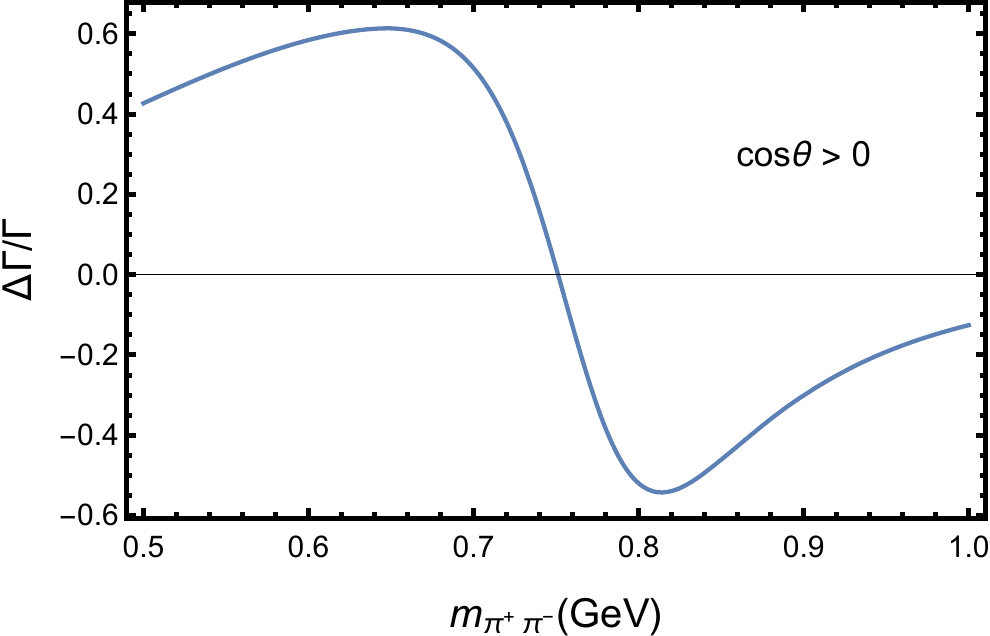}
}
\subfigure[]{
  \includegraphics[width=6.5cm]{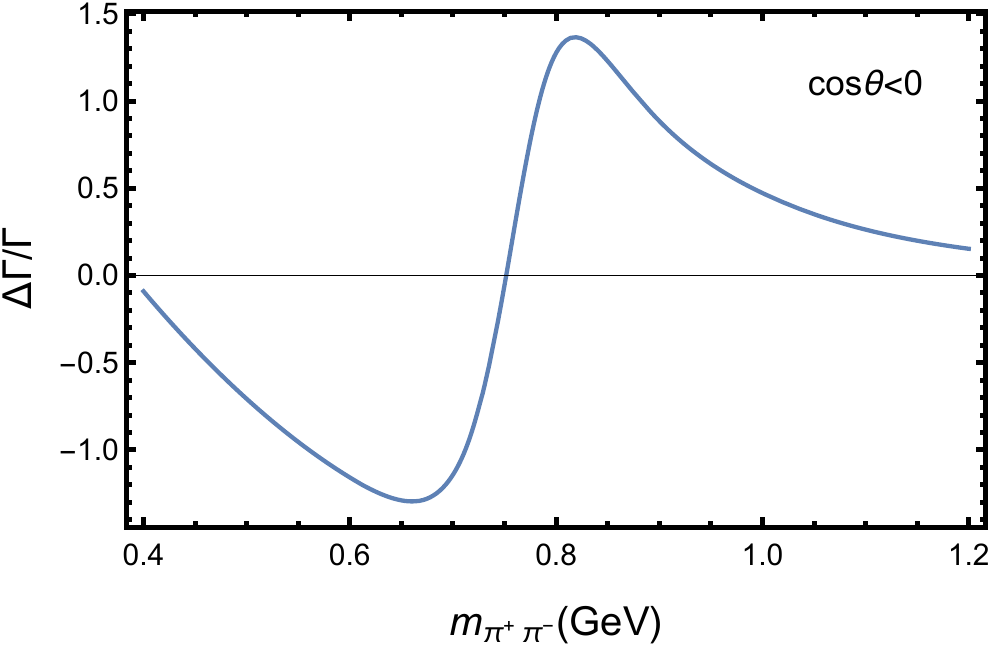}
}
\hspace{0.5cm}
\subfigure[]{
  \includegraphics[width=6.5cm]{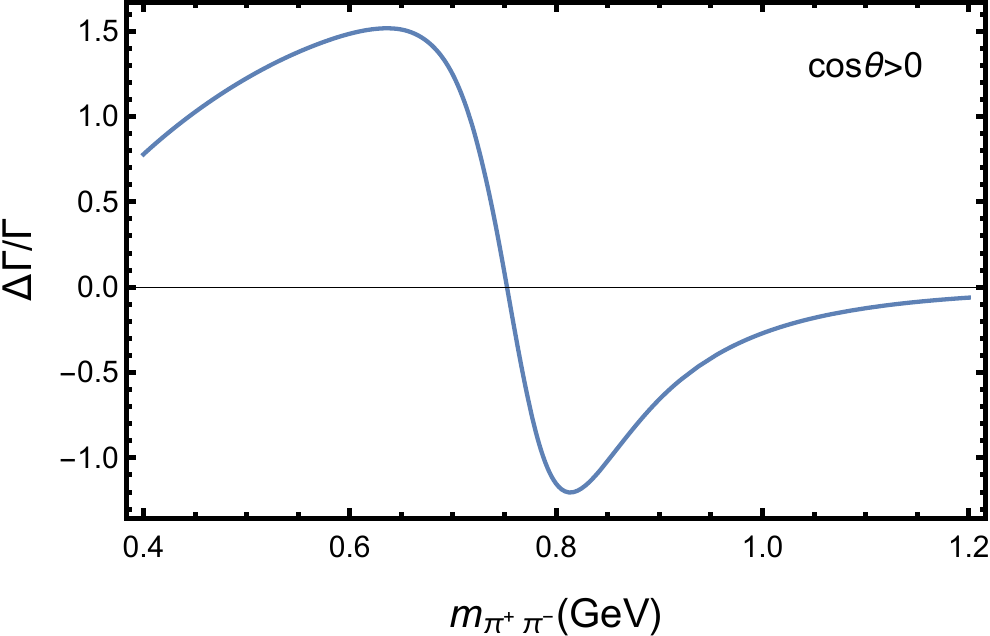}
}
\subfigure[]{
  \includegraphics[width=6.5cm]{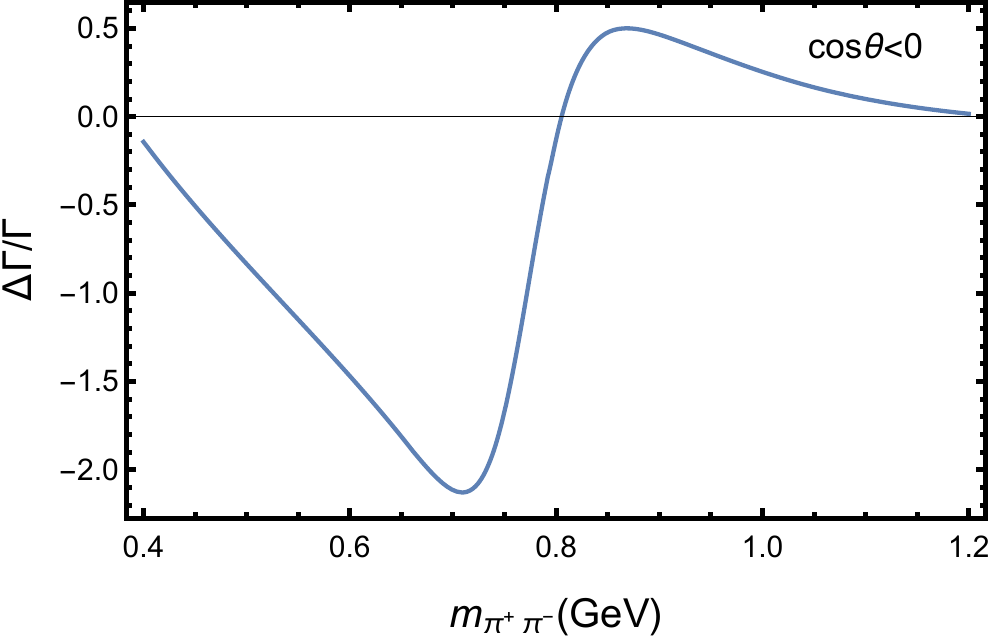}
}
\hspace{0.5cm}
\subfigure[]{
  \includegraphics[width=6.5cm]{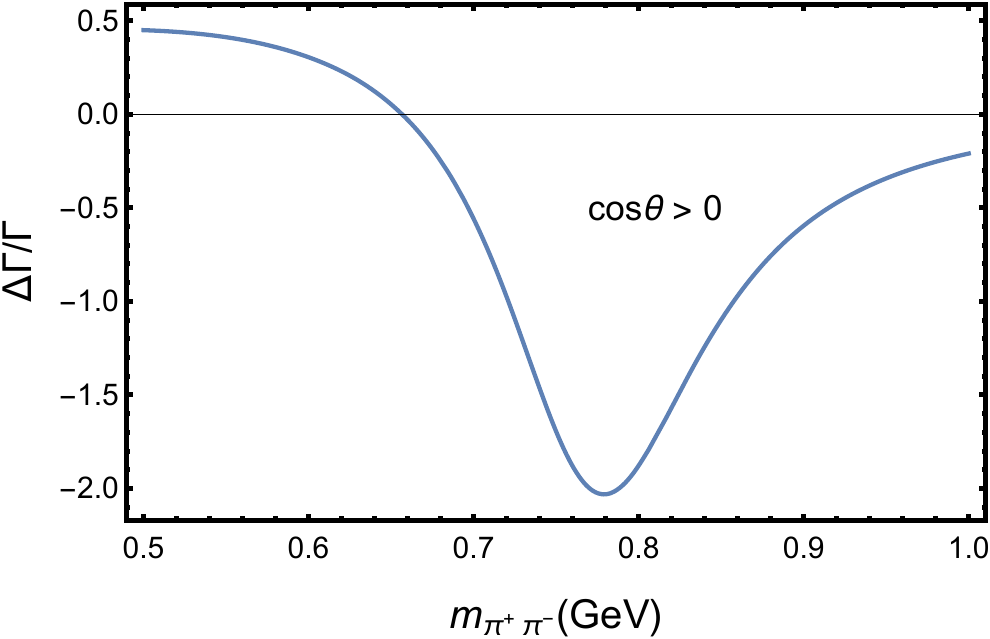}
}
\caption{\small The rate asymmetry $\Delta\Gamma$ in units of $\Gamma=1/\tau(B^\pm)$ for $B^\pm\to \pi^\pm\pi^+\pi^-$ in the low-$m_{\pi^+\pi^-}$ region induced by the interference between $\rho(770)$ and the $\sigma$ meson for (a) $\cos\theta<0$ or  $\cos\theta_{\rm hel}>0$ and (b) $\cos\theta>0$ or  $\cos\theta_{\rm hel}<0$. Rate asymmetries of $\rho(770)$ and $\sigma$ are added  to (a) and (b) and shown in (c) and (d), respectively. The phases of $A^\sigma_\pm$ relative to $A^\rho_\pm$ are given by Eq. (\ref{eq:phases}).
}
\label{fig: inter_1}
\end{figure}

This above-mentioned difficulty is resolved by noticing a nontrivial global phase difference between $\rho$ and $\sigma$. Indeed, the phase $\phi_-$ ($\phi_+$) of the $A^\sigma_\mp$ relative to $A^\rho_\mp$ in $B^-$ ($B^+$) decays has been measured by LHCb to be
\be
\phi_-^{\rm expt}=115\pm2\pm14\,, \qquad \phi_+^{\rm expt}=179\pm1\pm95\,,
\en
in units of degrees (see Table I of \cite{Aaij:3pi_1}).
Assigning the phases $\phi_-$ and $\phi_+$ to the amplitudes $A^\sigma_-$ and $A^\sigma_+$, respectively, we find the observed interference pattern in
Figs. \ref{fig:rho-sigma}(a) and \ref{fig:rho-sigma}(b) for $\cos\theta<0$ and $\cos\theta>0$, respectively, can be properly reproduced in
Figs. \ref{fig: inter_1}(a) and \ref{fig: inter_1}(b), respectively, with
\be \label{eq:phases}
\phi_-=115^\circ, \qquad\quad \phi_+=127^\circ.
\en
Since the measured phase $\phi_+$ has a large uncertainty, we have fixed $\phi_-$ to its central value $115^\circ$ and varied $\phi_+$ within the experimental allowed region. We find that the phase $\phi_+$ is in the vicinity of $127^\circ$. However, it is not clear to us what the dynamical origin of the phase difference is.

\begin{figure}[tbp]
\centering
\subfigure[]
{
  \includegraphics[width=4.9cm]{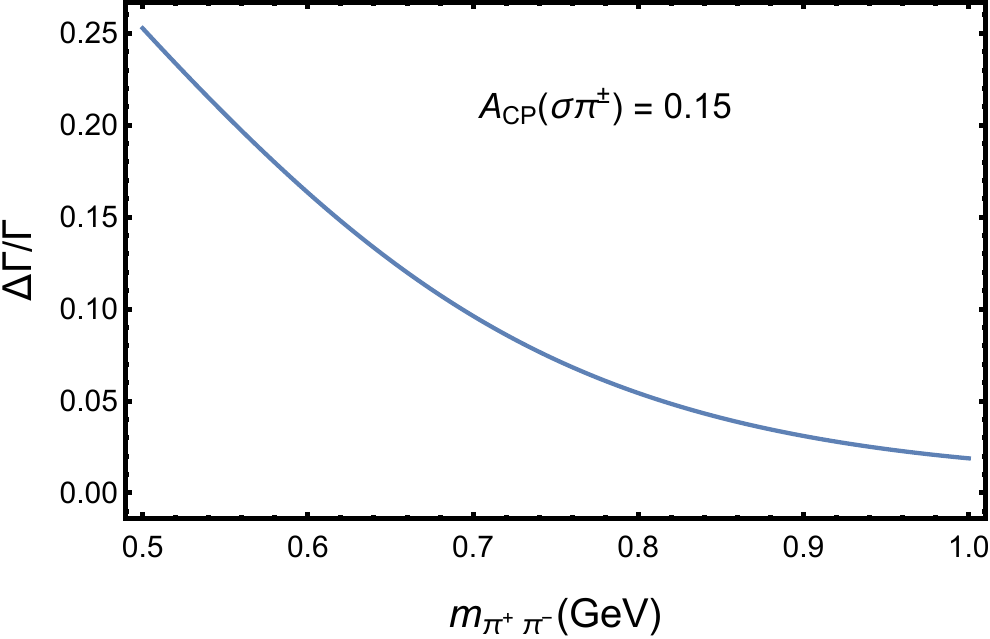}
}
\hspace{0.2cm}
\subfigure[]
{
  \includegraphics[width=4.9cm]{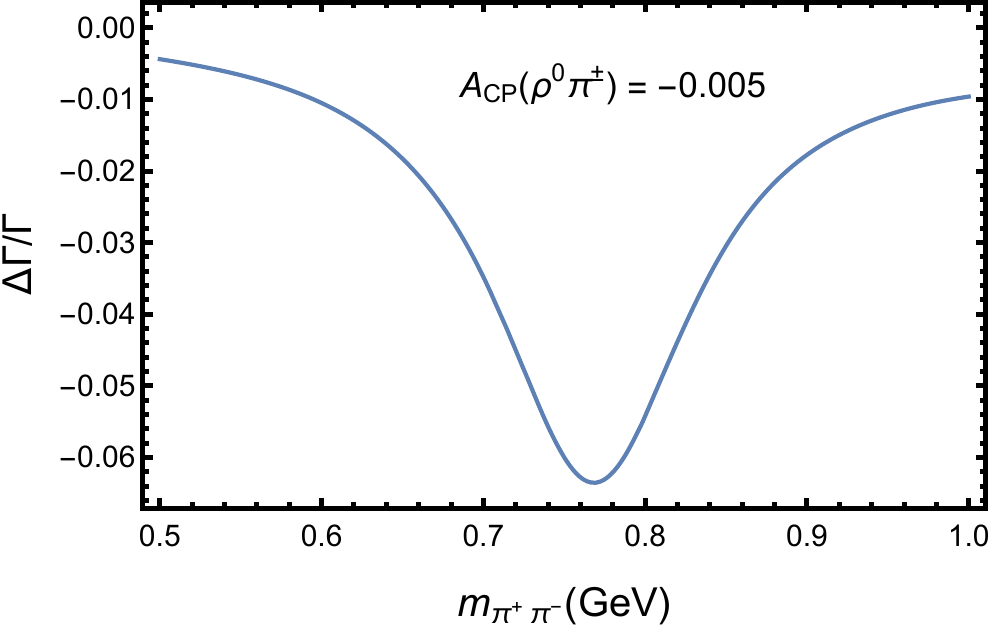}
}
\hspace{0.2cm}
\subfigure[]
{
  \includegraphics[width=4.9cm]{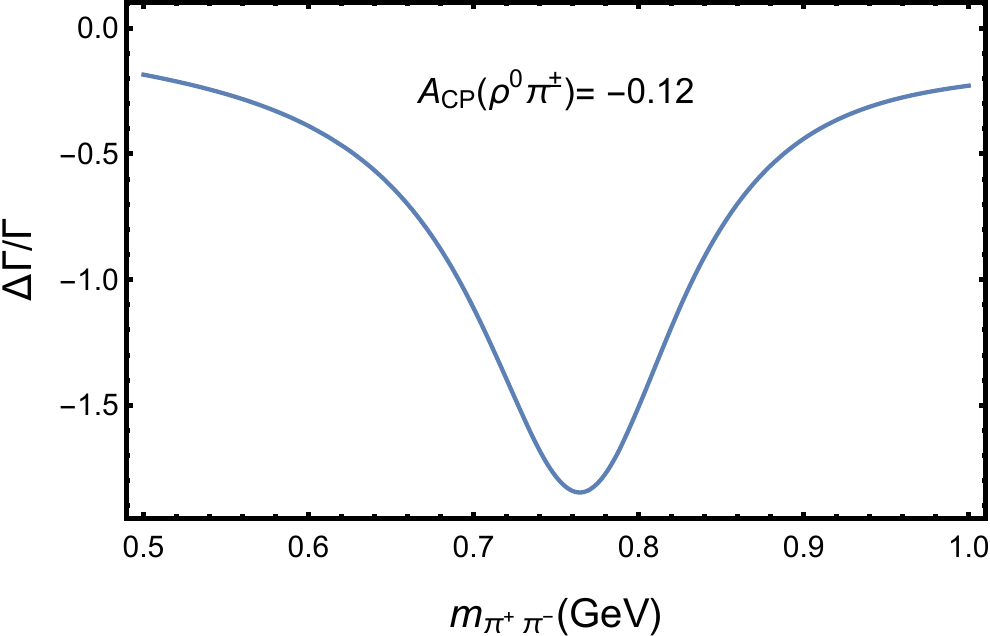}
}
\caption{\small Rate asymmetries $\Delta\Gamma$ in units of $\Gamma=1/\tau(B^\pm)$ for $B^\pm\to \pi^\pm\pi^+\pi^-$ in the low-$m_{\pi^+\pi^-}$ region induced by the $\sigma$ meson (a) and the $\rho(770)$ [(b) and (c)]  for either $\cos\theta<0$ or  $\cos\theta>0$. (a) corresponds to  $\A_{CP}
(B^\pm\to \sigma\pi^\pm)=0.15$\,,
 (b) to $\A_{CP}
(B^\pm\to \rho^0\pi^\pm)=-0.005$ and (c) to $\A_{CP}
(B^\pm\to \rho^0\pi^\pm)=-0.12$\,.
}
\label{fig: CPrho&sigma}
\end{figure}

Thus far we have neglected \CPP-violating contributions from $\rho$ and $\sigma$ to rate asymmetries, namely, the first two terms in Eq. (\ref{eq:modelCP}). Figs. \ref{fig: CPrho&sigma}(a) and \ref{fig: CPrho&sigma}(b) show the rate asymmetries of $B^\pm\to \sigma\pi^\pm\to \pi^+\pi^-\pi^\pm$ and $B^\pm\to \rho^0\pi^\pm\to \pi^+\pi^-\pi^\pm$, respectively, as a function of low-$m_{\pi^+\pi^-}$ between 0.5 and 1.0 GeV for either $\cos\theta<0$ or $\cos\theta>0$, \footnote{They do not change sign when the cosine varies from $-1$ to $+1$ as $|A^\rho_\pm|^2$ are proportional to $\cos\theta^2$, while $|A^\sigma_\pm|^2$ are $\theta$ independent.}
corresponding to  $\A_{CP} (B^\pm\to \sigma\pi^\pm)=0.15$ (see Eq. (\ref{eq:CPsigma})) and  $\A_{CP}
(B^\pm\to \rho^0\pi^\pm)=-0.005$ (see Table \ref{tab:rhopi_theory}).
After adding the rate asymmetries from $\rho$ and $\sigma$ from Figs. \ref{fig: CPrho&sigma}(a) and \ref{fig: CPrho&sigma}(b), respectively, to both Figs. \ref{fig: inter_1}(a) and \ref{fig: inter_1}(b), we finally get Figs. \ref{fig: inter_1}(c) and \ref{fig: inter_1}(d) for $\cos\theta<0$ and $\cos\theta>0$, respectively. By comparing them with Fig. \ref{fig:rho-sigma}, we see that the resultant rate asymmetries agree well with the experimental observation, namely the peak height and the bottom depth are similar in magnitude with the former slightly larger. The curve below $m_{\pi^+\pi^-}\sim 700$ MeV in  Fig. \ref{fig: inter_1}(d) becomes flat owing to the contribution from the $\sigma$ resonance at lower $m_{\pi^+\pi^-}$.

Two remarks are in order. (i) The real part of the GS line shape is proportional to $s-m_\rho^2-f(s)$ rather than $s-m_\rho^2$. Nevertheless, we find that the contribution from $f(s)$ is numerically small and can be neglected. (ii) rate asymmetries do not vanish exactly at $m_{\pi^+\pi^-}=m_\rho$ due to other contributions such as the real part of the line shape, see Eq. (\ref{eq:modelCP}).

If the \CP asymmetry of the quasi-two-body decay $B^\pm\to\rho^0\pi^\pm$ is not neligible, what will the rate asymmetry look like? For an illustration, we shall take the case (ii) in Table \ref{tab:rhopi_theory} as an example, namely, $\A_{CP}(B^\pm\to\rho^0\pi^\pm)=-0.12$ obtained in QCDF by setting  both
$\rho_H$ and $\phi_H$ to zero. $\Delta \Gamma/\Gamma$ due to the $\rho$ resonance is now shown in Fig. \ref{fig: CPrho&sigma}(c).
The rate asymmetries due to the $\rho-\sigma$ interference
are very similar to that of Figs. \ref{fig: inter_1}(a) and \ref{fig: inter_1}(b) for $\cos\theta<0$ and $\cos\theta>0$, respectively. Adding contributions from Figs. Fig. \ref{fig: CPrho&sigma}(a) and \ref{fig: CPrho&sigma}(c) to Figs. \ref{fig: inter_1}(a) and \ref{fig: inter_1}(b) yield the final rate asymmetries in Figs. \ref{fig: inter_1}(e) and \ref{fig: inter_1}(f) for $\cos\theta<0$ and $\cos\theta>0$, respectively.  It is obvious that \CP asymmetries are pushed downward owing to the negative \CP violation in $B^\pm\to \rho^0\pi^\pm$. Evidently, the patterns exhibited in  Figs. \ref{fig: inter_1}(e) and \ref{fig: inter_1}(f) are not consistent with the data. Indeed, rate asymmetries do not cancel when integrating over the angle.
We thus conclude that the experimental observation of the interference pattern between $P$- and $S$-waves in the low-$m_{\pi^+\pi^-}$ region between 0.5 and 1.0 GeV is consistent with a nearly vanishing \CP violation in $B^\pm\to\rho^0\pi^\pm$.

For the interference between $P$- and $S$-waves, we have thus far focused on the $S$-wave contribution from the $\sigma$ resonance. In principle, the interference of $\rho(770)$ with $f_0(980)$ and the nonresonant component as considered in \cite{Bediaga:2015} is also allowed. Since the quasi-two-body decay $B^+\to f_0(980)\pi^+$ is quite suppressed \cite{Cheng:2013}, the $f_0(980)$ fraction should be less than 1\% (see e.g. Table VI of Ref. \cite{Cheng:2020ipp}), much smaller than the $\sigma$ fraction of 25\% in the isobar model. The nonresonant contribution was not specified in the LHCb analysis, though it has been studied in the earlier BaBar measurement \cite{BaBarpipipi}. Therefore, in this work we will not consider the interference of $\rho(770)$ with $f_0(980)$ and the nonresonant background.

\section{Conclusions}
The decay amplitudes of $B^\pm\to \pi^+ \pi^-\pi^\pm$  in the Dalitz plot analyzed by the LHCb indicate that \CP asymmetry for the dominant quasi-two-body decay $B^\pm\to\rho(770)^0\pi^\pm$ was found to be consistent with zero in all three approaches for the $S$-wave component and that  \CPP-violation effects related to the interference between the $\rho(770)^0$ resonance and the $S$-wave were clearly observed. We show that the nearly vanishing \CP violation in $B^\pm\to\rho^0\pi^\pm$ is understandable in the framework of QCD factorization. There exist two $1/m_b$ power corrections, one from penguin annihilations and the other from hard spectator interactions.
In the heavy quark limit, $\acp(B^\pm\to\rho^0\pi^\pm)$ turns out to be positive. Hard spectator interactions will push it up further, whereas penguin annihilation will pull it to the opposite direction. These two power corrections contribute destructively to $\acp(B^\pm\to\rho^0\pi^\pm)$ to render it consistent with zero, in contrast to the non-negligible \CP asymmetry predicted in most of the existing models.

We next show that the measured interference pattern between the $\rho$ and $S$-wave contributions in the low $m_{\pi^+\pi^-}$ region separated by the sign of value of $\cos\theta$  can be explained in terms of the smallness of $\acp(B^+\to\rho^0\pi^+)$ and the interference between $\rho(770)$ and $\sigma/f_0(500)$. There are two key ingredients needed for producing the observed interference curve: one is the global phase difference between $\rho(770)$ and $\sigma$ contributions in $B^-$ and $B^+$ decays, which has been measured by LHCb. The other is the long-distance strong phase arising from the line shape of the intermediate resonance. In particular, the real and imaginary parts of the Breit-Wigner propagator with the $\rho$ resonance are proportional to $s-m^2_\rho$ and $m_\rho\Gamma_\rho$, respectively. The characteristic feature of the interference pattern, namely, \CP asymmetry is proportional to $(s-m^2_\rho)\cos\theta$, is thus accounted for, as shown in Figs. \ref{fig: inter_1}(c) and \ref{fig: inter_1}(d).
If \CP asymmetry in $B^\pm\to\rho^0\pi^\pm$ is not negligible as predicted in many existing models, the observed interference pattern will be destroyed, see e.g. Figs. \ref{fig: inter_1}(e) and \ref{fig: inter_1}(f). This again implies a nearly vanishing \CP violation in $B^\pm\to\rho^0\pi^\pm$.

\vskip 2.0cm \acknowledgments
This research was supported in part by the National Science and Technology Council of R.O.C. under Grant No. 111-2112-M-001-036.


\end{document}